\documentclass[reprint,amsmath,amssymb,aps,twocolumn]{revtex4-2}
\usepackage{graphicx}
\usepackage{dcolumn}
\usepackage{rotating}
\usepackage{amsthm,amsmath,amssymb,amsfonts,bbm}
\usepackage{mathptmx}
\usepackage{mathrsfs}
\usepackage{siunitx}
\usepackage{bm}
\usepackage{textcomp}
 
\begin{document}

\title{Image-charge detection of electrons on helium in an on-chip trapping device}

\author{Mikhail Belianchikov$^{1,*}$, Natalia Morais$^{1}$, Denis Konstantinov$^{1}$}
\affiliation{$^{1}$Quantum Dynamics Unit, Okinawa Institute of Science and Technology (OIST) Graduate University, Onna, 904-0495 Okinawa, Japan}
\email[E-mail:]{m.belianchikov@oist.jp}

\date{\today}

\begin{abstract}
Electrons trapped on the surface of superfluid helium have been thought of as a useful resource for quantum computing. Such electrons show long coherence of their surface-bound (Rydberg) states combined with their easy electrostatic manipulation. Recent proposals explored the possibility of coupling the spin state of an electron and the state of its quantized motion with a promise of a highly scalable 2D architecture for a quantum computer. However, despite recent progress in the detection of quantized lateral motion of electrons using a circuit-QED setup, the manipulation of a small number of electrons and their quantum state detection remains a challenging problem. Here, we report on the detection of the Rydberg transition of electrons on superfluid helium in an on-chip microchannel device in which electrons are moved and trapped by a set of electrostatic gates. A highly sensitive image-charge detection system allows us not only to resolve the transition spectra of such electrons, but also to perform the device characterization. The demonstrated sensitivity shows the feasibility of detecting the Rydberg transition of a single electron, which can open a new pathway for a non-destructive spin-state readout. 
\end{abstract}

\maketitle

\section{Introduction}
\label{sec:intro}

Electrons bound to the surface of liquid helium have been considered to be a promising resource for quantum computing~\cite{platzman1999qubits,Dykman2003,lea2000qubits,Lyon2006,Schuster2010}. The surface-bound states (also referred to as the Rydberg states~\cite{CollinPRL2002}) of such electrons are formed due to a long-range polarization attraction to the dielectric substrate and a short-range repulsion from the helium atoms~\cite{Cole1970-PRB}. Thanks to the pristine host environment free of impurities and defects, the system is known for its very high mobility~\cite{dahmPRB1987,shirahamaJLTP1996mobility,Andrei} and predicted spin coherence time exceeding 100 seconds~\cite{Lyon2006}. Significant recent progress has been achieved by trapping electrons and detecting the quantized states of their lateral motion using a circuit-QED setup~\cite{YangPRX2016,Koolstra2019}. A remarkable result was achieved by employing a similar method for electrons trapped on solid neon~\cite{zhou2022_Nature}, demonstrating the coherence time of lateral motion states on the order 0.1 millisecond~\cite{zhou2023_NatPhys}. Moreover, a recent theoretical study suggests the coupling between the spin of an electron to its Rydberg states as a new pathway towards building a scalable architecture of high-fidelity spin-qubit gates~\cite{kawakamiPRA2023qubits,kawakamiAPL2024review}. While this is a promising approach toward a fault-tolerant quantum computer consisting of millions of qubits, it is still very challenging to realize the spin-state readout and manipulation of electrons on such a large scale. It was proposed to detect the spin state of a coupled Rydberg-spin qubit nondestructively by detecting its Rydberg transition~\cite{kawakamiPRA2023qubits}. Since the typical Rydberg transition frequency is on the order of hundreds of GHz, the developed methods of the circuit-QED based on a superconducting coplanar waveguide (CPW) resonator are not suitable for such a high frequency. Thus, a new method must be developed to detect the Rydberg transition of a single electron to realize the single-spin readout and spin-qubit logic gates. Moreover, a new experimental platform has to be sought to realize the manipulation of a large number of electrons towards building a scalable architecture of logic gates.

A new method of the image-charge detection has been recently realized to detect the Rydberg transition in a large ensemble ($\sim 10^8$) of electrons on liquid helium~\cite{kawakamiPRL2019image}. The method exploits an occurrence of the image current induced by the excited electrons in a metal electrode placed close to the electron system. Despite its conceptual simplicity, this method proved to be a valuable tool for studying nontrivial dynamics of the photo-excited electrons on liquid helium, such as the excited-state relaxation due to interaction with surface vibrations (ripplons)~\cite{kawakami2021relaxation} and the thermoelectric (Seebeck) effect in a correlated electron system~\cite{KostylevPRL2002}. A recent experiment with electrons confined in an array of 4~\textmu m deep channels filled with superfluid helium also suggested a route towards improving the sensitivity of the image-current detection and a possibility to use such a setup for scaling the detection method towards a single electron readout~\cite{zouNJP2022image}. The devices comprised of microchannels filled with superfluid helium (also referred to as the microchannel devices) have been used to study the transport properties of surface-bound electrons for decades~\cite{glassonPRL2001fet,ikegamiPRL2009wigner,reesPRL2011pc,reesPRL2016slip,badrutdinovPRL2020polaron,GuoMeJin}. In such devices, the number of electrons confined within a channel can be electrostatically manipulated with the gate electrodes in a manner reminiscent of a semiconductor-based field-effect transistor (FET)~\cite{klierJLTP2000fet}. Electrons can be also transferred along the channels in a manner of a classical charge-coupled device (CCD), demonstrating an unprecedented efficiency of 99.99999999\%~\cite{bradburyPRL20011ccd}. Such an exquisite purely electrostatic control over the system could be very useful to build a large-scale quantum computer, providing that a robust control and readout of electronic quantum states is established~\cite{Pino2021,haffPRA2022,lanPRX2023}.  

In this work, we demonstrate the implementation of a microchannel FET-like trapping device with an image-charge detection system to detect the Rydberg transition in a small ensemble of electrons trapped and manipulated in a single 10-\textmu m wide and 100-\textmu m long channel trap. We find that the evolution of the transition Stark spectra with applied gate voltages is described well even by a simple analytical model of a uniformly charged helium surface. The observed spectra also reveal some unexpected details about the charge depletion and trapping inside the channel, compared with what is usually learned from conventional transport measurements. An apparent inconsistency with the current-voltage characteristics of the device is clarified by the finite-element modeling (FEM) of the trapping potential in the channel. Thanks to the very high sensitivity of our detection system based on a superconducting helical resonator, we could observe the transition signal from a few excited electrons, with the total number of electrons in the trap estimated to be on the order of hundreds. The demonstrated sensitivity of our image-charge detection system shows the feasibility of detecting the Rydberg transition of a single electron, which would open a pathway for single-spin detection and realizing spin-qubit gates in a trapping device. Together with the microchannel device capabilities of the precise control of electrons and seamless integration of on-chip devices with CMOS technology, this can open a new pathway for realizing a large-scale quantum computer. 
 
\section{Experimental details}
\label{sec:setup}

\subsection{Setup and methods}

\begin{figure*}[htp]
\includegraphics[width=\textwidth,keepaspectratio]{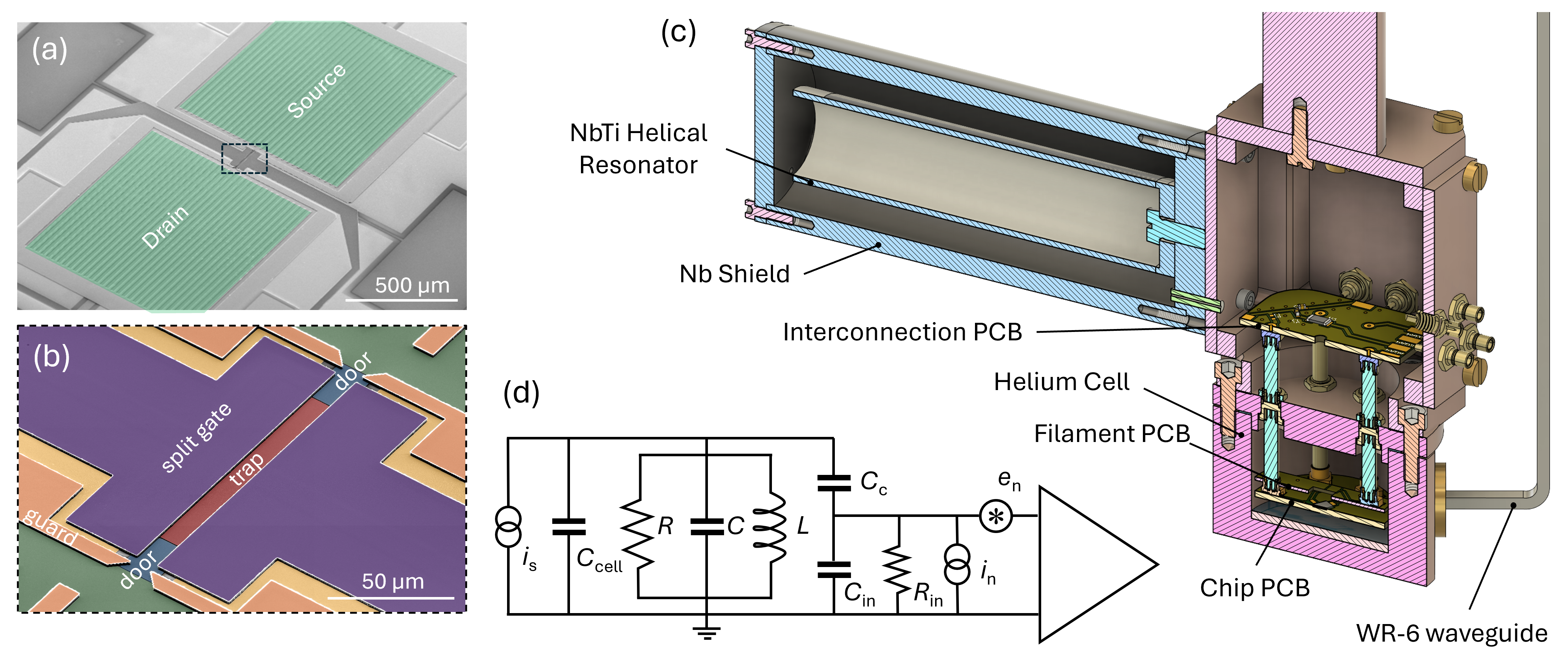} 
\caption{\label{fig:1} (color on line) Experimental setup. (a) False-color micrograph of the microchannel device for electron storage and manipulation. Two parallel-channel arrays (green) serving as the source and drain of FET are connected by a central channel (enclosed in dashed rectangular) where the number of electrons can be varied by the DC bias voltages applied to the gate electrodes. (b) Zoomed image of the central channel showing the gate electrode structure comprised of three electrodes (door/trap /door) at the bottom layer and two split-gate electrodes (on two sides of the channel) at the top layer. (c) 3D schematic image of the cryogenic image-charge detection setup. (d) Equivalent electrical circuit of the detection system including signal and noise sources, as described in the text.}
\end{figure*}

The experimental setup consists of a microchannel trapping device for electrons manipulation and an image-charge detection system for their sensing (see Fig.~\ref{fig:1}). Our microchannel trapping device shown in Fig.~\ref{fig:1}(a) has a standard FET source/gate/drain structure comprised of two identical arrays of 20~\textmu m wide and 700~\textmu m long channels, which serve as two reservoirs (source and drain) for electron storage, connected by a single 10~\textmu m wide and 140~\textmu m long central channel (gate) where the electron number can be electrostatically controlled. The structure is fabricated on a high-resistive silicon chip using maskless UV lithography. It comprises two gold layers of 50 nm thickness and a 1.5~\textmu m-thick insulator layer between them. The bottom gold layer consists of biasing electrodes, which define the bottom of the reservoirs and the central channel. A distinctive feature of our trapping device is a three-electrode structure of the central channel bottom layer, which consists of two 22~\textmu m long {\it door} electrodes separated by a 100~\textmu m long {\it trap} gate electrode (see Fig.~\ref{fig:1}(b)). In the experiment described here, the two door electrodes are connected to form a door gate. The top layer defines the channel structure and consists of two electrodes, {\it guard} and {\it split gate}, to confine electrons in the reservoirs and central channel, respectively. The insulator layer, made of hard-baked photoresist, is etched down to the bottom metal layer using the top metal layer as a hard mask. The chip is mounted on a FR4 printed-circuit board (PCB) inside a leak-tight experimental cell attached to the mixing chamber of a dilution refrigerator and partially filled with liquid helium (see Fig.~\ref{fig:1}(c)).  

The image-charge detection setup is an improved version of the previously reported~\cite{mishaJLTP2023amp}. At the heart of the method is a cryogenic image-current amplifier represented by a superconducting helical resonator. Our resonator is comprised of a coil inductor with inductance $L=4.15$~mH made of copper-clad free NbTi wire (bare diameter 50~\textmu m) with PTFE insulation (Supercon Inc.) inclosed in a Nb shield (see Fig.~\ref{fig:1}(c)). One end of the inductor is AC grounded, while the other is connected to the DC-biased trap gate of the microchannel device inside the cell. Together with the self-capacitance of the coil and parasitic capacitance of the cell, the inductor forms a parallel RLC (tank) circuit resonating at the frequency $f_0\sim 1$~MHz. The high impedance of the tank circuit at resonance $R$ is used to convert the image-current signal $i_\textrm{s}$ induced at the trap gate of the microchannel device by the photo-excited electrons to a voltage signal $i_\textrm{s}R$, which is capacitively coupled to a cryogenic low-noise amplifier (LNA) (CX-4, Stahl-Electronics). To resonantly detect the image current of electrons induced by the photo-excited electrons above the trap gate, the millimeter-wave (mm-wave) excitation from a room-temperature source transmitted into the cell via a waveguide is pulsed-modulated (PM) at the frequency $f_\textrm{m}=f_0$ (the duty cycle of 15\% was used to reduce heating of the cell by the radiation). The voltage signal at the LNA output is detected using a lock-in amplifier referenced at the modulation frequency $f_\textrm{m}$. The lock-in input senses the image-charge signal due to electrons exposed to PM mm-wave excitation, confirmed by observing zero signal when all electrons are removed from the device.  

\subsection{Noise and sensitivity}
\label{sec:noise}

To estimate the sensitivity of our detection method, it is important to account for all sources of noise in the system. The equivalent circuit of our image-charge detector is shown in Fig.~\ref{fig:1}(d). The image-charge signal is represented by an ideal current source $i_\textrm{s}$ in parallel with the cell capacitance $C_\textrm{cell}$ and the superconducting tank circuit RLC. The latter is connected via a coupling capacitance $C_\textrm{c}=1$~pF to the amplifier with input resistance $R_\textrm{in}$ and input capacitance $C_\textrm{in}$. The input resistance $R_\textrm{in}>40$~$\textrm{M}\Omega$ contributes negligibly to the circuit. Therefore, it will be omitted from further consideration. According to the standard model~\cite{NoiseConnely}, the internal sources of noise at the input of LNA can be represented by an ideal voltage source $e_n$ and an ideal current source $i_n$ connected to the LNA input in series and in parallel, respectively. Together with the thermal noise of the tank circuit at resonance, the spectral density of the total voltage noise at the input of an ideal amplifier in Fig.~\ref{fig:1}(d) can be written as~\cite{jeffRSI1993}

\begin{equation}
e_\textrm{in}^2=4k_\textrm{B}T R \kappa^2 + i_n^2 R^2 \kappa^4 +e_n^2,
\label{eq:noise}
\end{equation}   

\noindent where $T$ is the temperature of the tank circuit, $k_\textrm{B}$ is the Boltzmann constant, and $\kappa = C_\textrm{c}/(C_\textrm{c}+C_\textrm{in})$. Note that the above expression is valid only in a narrow frequency range near the resonance frequency of the tank circuit given by $f_0 = (2\pi\sqrt{L(C+C_\textrm{cell}+C_\textrm{c}||C_\textrm{in})})^{-1}$, where the capacitance $C_\textrm{c}||C_\textrm{in}$ is given by the series combination of $C_\textrm{c}$ and $C_\textrm{in}$. 

\begin{figure}[htp]  
\includegraphics[width=\columnwidth,keepaspectratio]{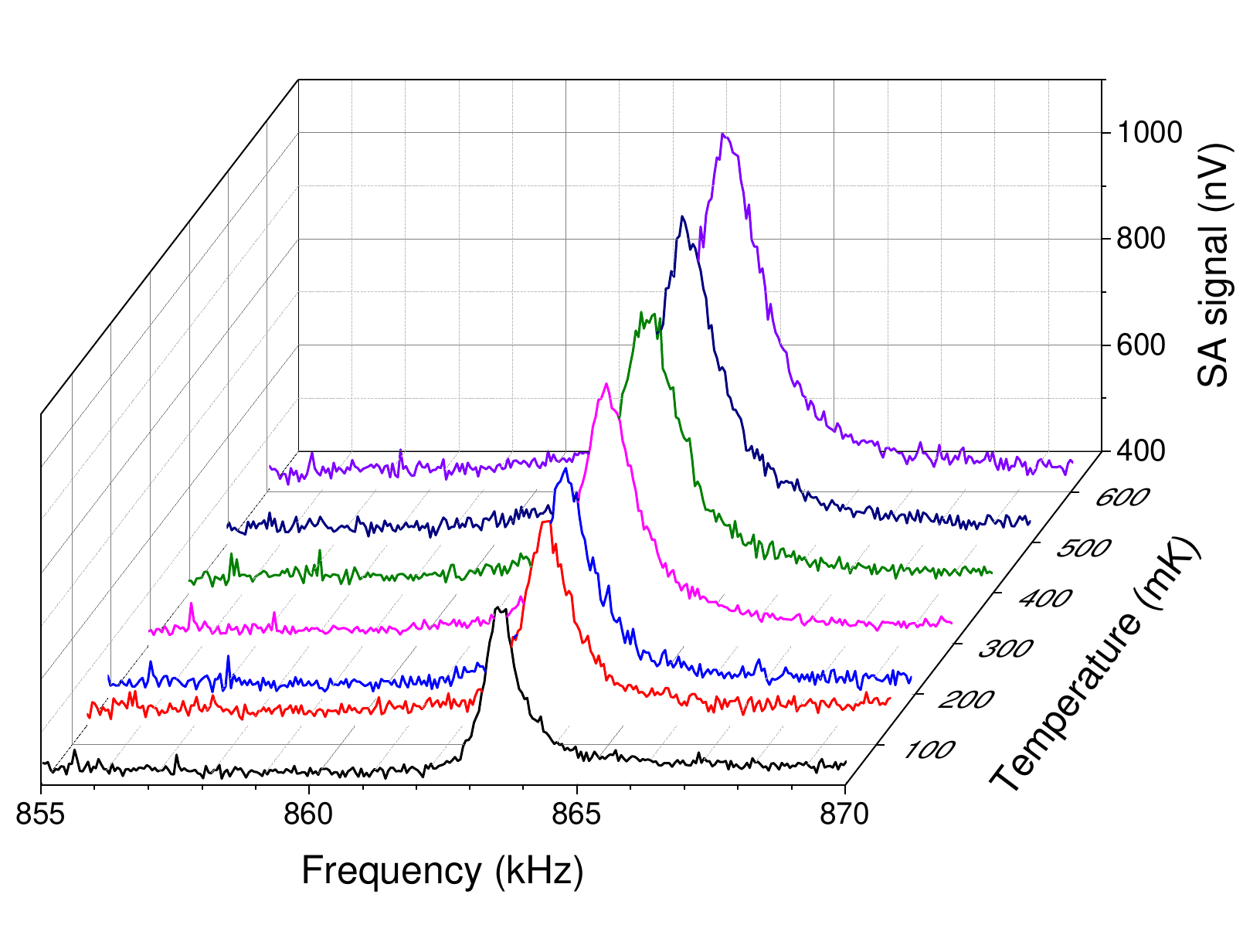} 
\caption{\label{fig:1app} (color on line) Noise spectra of the tank circuit measured at different temperatures at the output of the amplifier using a spectrum analyzer.}
\end{figure}

\begin{figure}[htp]  
\includegraphics[width=\columnwidth,keepaspectratio]{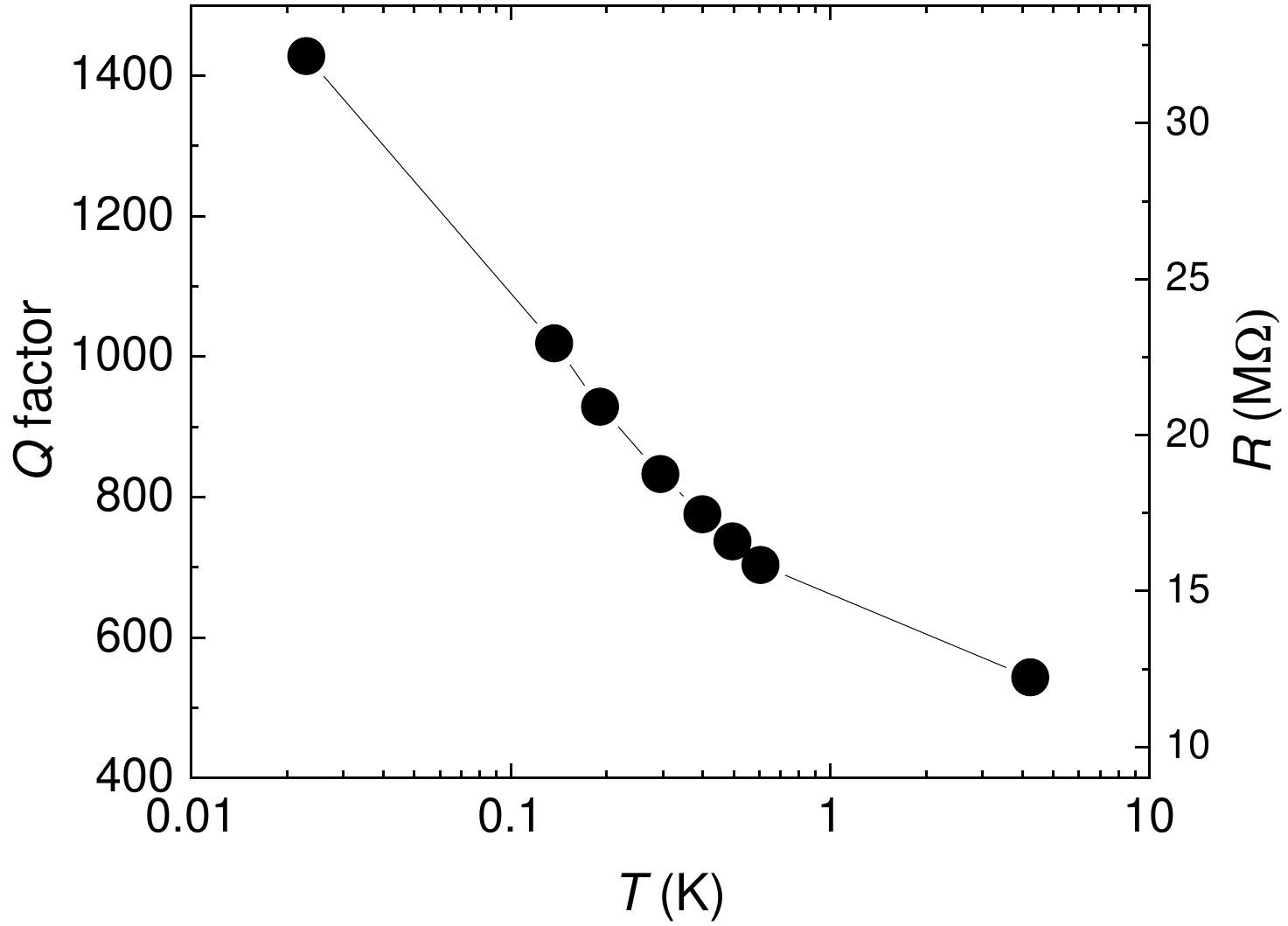} 
\caption{\label{fig:2app} (color on line) Loaded quality factor of the tank circuit obtained from the noise spectra shown in Fig.~\ref{fig:1app}.}
\end{figure}

\begin{figure}[htp]
\includegraphics[width=8cm,keepaspectratio]{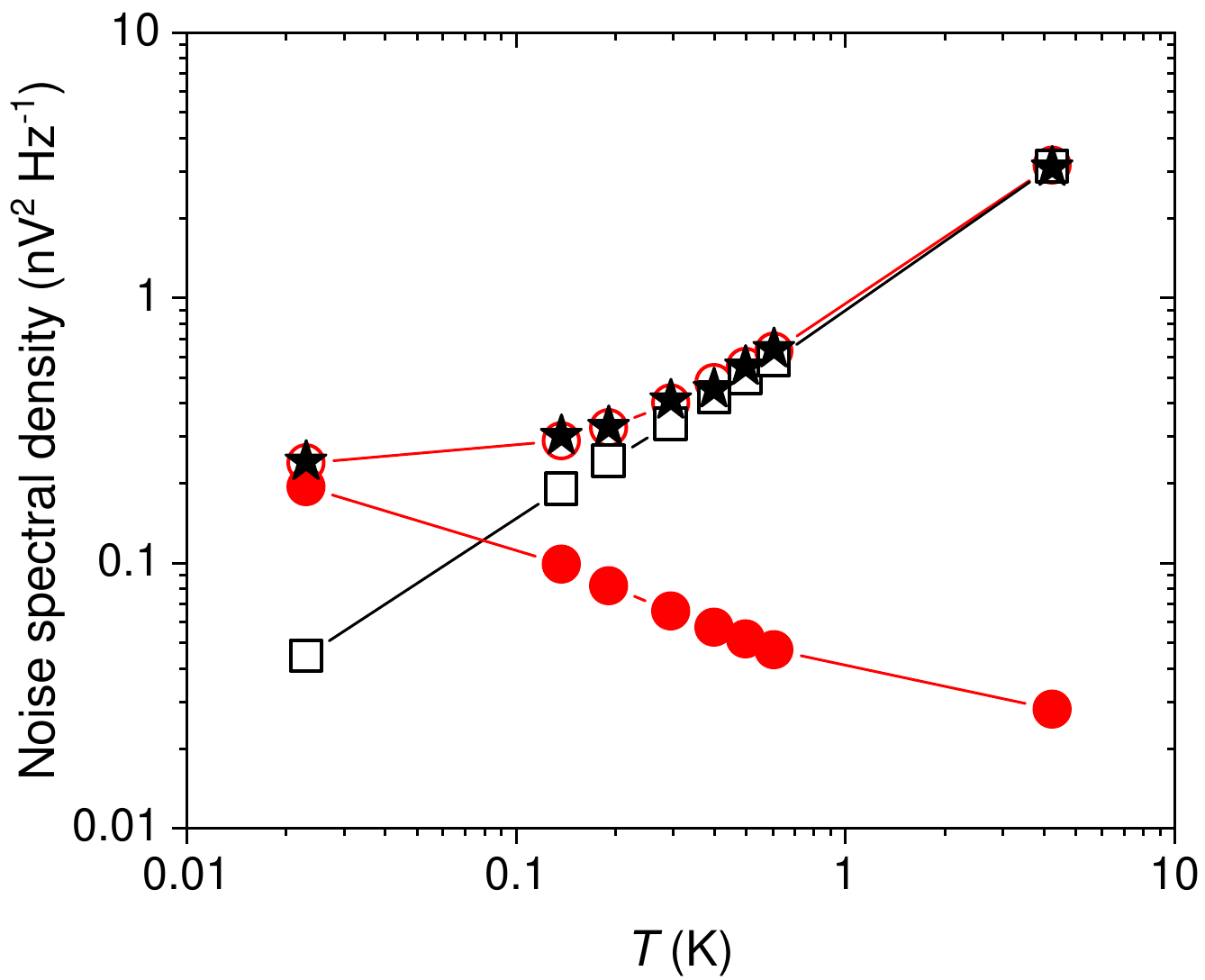} 
\caption{\label{fig:2} (color on line) Temperature-dependent contribution to the spectral density of voltage noise at the amplifier input obtained from the measured noise spectra (star symbols) and calculated using Eq.~\eqref{eq:noise} (open circles). The calculated contribution due to the thermal noise of the tank circuit and the noise current of the amplifier are plotted by open squares and solid circles, respectively.}
\end{figure}

The above noise model provides a very convenient way to estimate different parameters of our detection system using the tank circuit near its resonance as a calibrated noise source. For this purpose, we measure the noise spectra at the amplifier output for different temperatures $T$ using a spectrum analyzer (SA). The results are shown in Fig.~\ref{fig:1app}. The RMS noise voltage at the amplifier output is given by $e_\textrm{in}A\sqrt{B}$, where $A$ is the voltage gain coefficient of LNA and $B$ is the resolution bandwidth of SA. From the measured spectra, we can determine both the resonance frequency of the tank circuit $f_0=863.55$~kHz, which is independent of temperature, and the loaded quality factor $Q$, which has a strong T dependence shown in Fig.~\ref{fig:2app}. Correspondingly, the impedance of the tank circuit at resonance given by $R=2\pi f_0 LQ$ also depends on $T$. The noise floor of the recorded spectra was independent of both the temperature and frequency in the considered range; thus, it can be attributed to the voltage noise of the amplifier $e_n$. Using $A=1,200$, which is provided by the manufacturer (Stahl-Electronics), and the resolution bandwidth of SA, which was set at $B=1$~Hz, we obtain $e_n=0.35$~nV/$\sqrt{\rm Hz}$. This agrees well with the datasheet value $e_n=0.33$~nV/$\sqrt{\rm Hz}$. The temperature-dependent contribution to $e_\textrm{in}^2$ in Eq.~\eqref{eq:noise} is obtained by subtracting $e_n^2$ and is plotted in Fig.~\ref{fig:2} by star symbols. This plot is obtained by considering the peak value of the measured noise voltage at the resonance frequency $f_0$. In order to account for the observed temperature dependence, which stems from the thermal noise and the temperature dependence of $R$, we plot the calculated thermal noise spectral density $4k_\textrm{B}TR\kappa^2$ and voltage noise spectral density $i_n^2R^2\kappa^4$ in Fig.~\ref{fig:2} by opened squares and solid circles, respectively. We consider the amplifier noise current $i_n$ and the coupling coefficient $\kappa$ to be unknown, and their values are adjusted such as the sum of these two contributions (plotted by opened circles in Fig.~\ref{fig:2}) match the measured values. From this fitting we find $\kappa=0.033$ and $i_n=12.5$~fA/$\sqrt{\textrm{Hz}}$. The latter is in good agreement with the datasheet value $i_n=11~$fA/$\sqrt{\textrm{Hz}}$. The estimated value of $\kappa$ suggests the input capacitance $C_\textrm{in}=29.2$~pF, which comes from the specified input capacitance of the amplifier 4.2~pF, the estimated capacitance of a connecting coaxial cable 17.4~pF, and stray capacitance on the interconnection PCB.

Finally, knowing the coupling coefficient $\kappa$, we can estimate the sensitivity of our detection system. The current signal $i_s$ produces the voltage signal $i_sR\kappa$ at the amplifier input (see Fig.~\ref{fig:1}(d)). The corresponding signal-to-noise ratio (S/N) is given by $\textrm{S/N}=i_sR\kappa/\sqrt{e_\textrm{in}^2B}$, with $e_\textrm{in}^2$ is given by Eq.~\eqref{eq:noise}. For example, with the measured value of $e_\textrm{in}^2=0.36$~$\textrm{nV}^2$/Hz at $T=23$~mK the signal-to-noise ratio S/N=1 corresponds to the current signal $i_s=0.57$~fA measured with the bandwidth of 1~Hz. Given the distance between the helium surface and the trap electrode $d=1.5$~$\mu$m, we can estimate the corresponding image-current signal due to the Rydberg transition of an electron, induced by PM mm-wave excitation at the modulation frequency $f_\textrm{m}=863.55$~kHz as $\pi e f_\textrm{m}  P_\textrm{e} (\Delta z /d)$, where $\Delta z\approx 10$~nm is the difference between the mean values of the vertical coordinate of electron occupying the ground state and the first excited Rydberg state and $P_\textrm{e}$ is the probability to occupy the first excited state~\cite{kawakamiPRL2019image}. Assuming $P_\textrm{e}=1$, we obtain $i_s\approx 2.9$~fA and, using the above noise estimate, we obtain signal-to-noise ratio $\textrm{S/N}=5$ for the detection of a single-electron transition with the measurement bandwidth of 1~Hz. In reality, the occupation probability $P_\textrm{e}$ depends on other experimental parameters. For the measured image-current response averaged over many microwave pulses $P_\textrm{e}$ is given by the matrix element $\rho_{22}$ of the reduced density operator of the system. Its dynamics are governed by the optical Bloch equations and explicitly depend on the Rabi frequency of mm-wave excitation and coherence rate of the Rydberg states due to the interaction of electrons with the ripplon bath. Unfortunately, the Rabi frequency can not be precisely measured in our experiment, and only its estimation from the measured input mm-wave power can be made. This is discussed further in Sec.~\ref{sec:closed}.  

\subsection{Device characterization}
\label{sec:device}

\begin{figure}[htp]
\includegraphics[width=7.5cm,keepaspectratio]{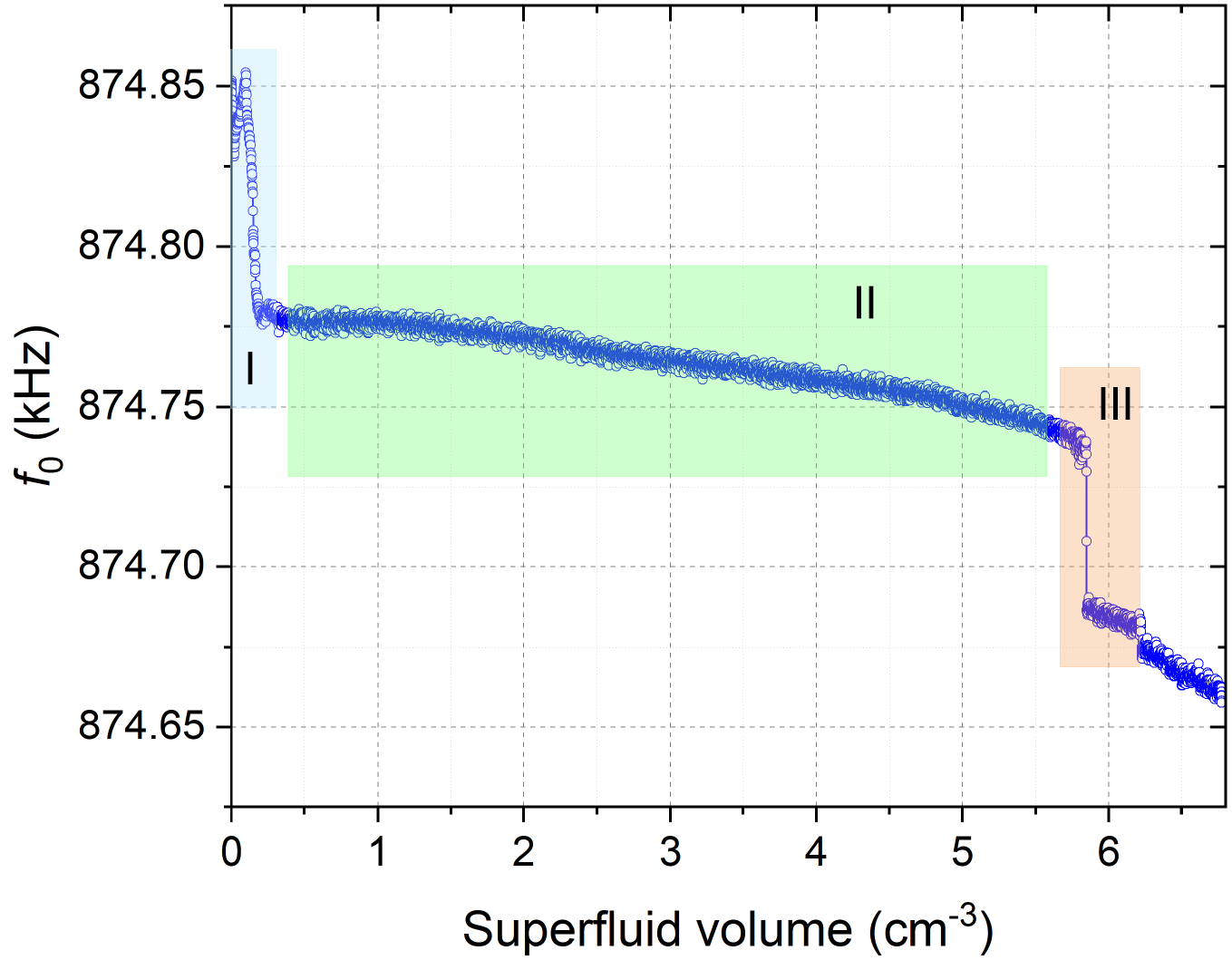} 
\caption{\label{fig:3} (color on line) Tank-circuit response to superfluid helium inside the channels showing the dependence of the resonant frequency on the amount of helium admitted into the cell. Different regimes of channel filling (I-III) are discussed in the text.}
\end{figure}

Reliable control of the superfluid helium thickness inside the channels is an important prerequisite for successfully trapping and manipulating electrons. In the experiment, the condensed helium gas admitted into the cell through a capillary line fills the bottom part of the cell below the chip, while the interior of the cell becomes coated with the superfluid film due to the van der Waals interaction, the film thickness being determined by the height $H$ with respect to the bulk liquid. In our setup, the filling of the microchannel device can be accurately monitored by observing the downshift of the resonant frequency of the tank circuit $f_0$, which is sensitive to the change in capacitance due to the dielectric liquid inside the channels. Fig.~\ref{fig:3} shows the circuit's resonant frequency versus the volume of superfluid helium inside the cell calculated from the amount of He gas admitted into the cell. Three distinct regimes, I, II, and III (highlighted by color), can be observed. With a small amount of liquid condensed inside the cell, a thin, unsaturated van der Waals film evenly coats the channel walls (regime I in Fig.~\ref{fig:3}), and the resonance frequency remains high. An abrupt drop occurs when the film saturates and starts filling the groves by the capillary action. A concave meniscus of liquid with the radius $R=\sigma/\rho g H$, where $\rho$ and $\sigma$ are the density and surface tension of liquid helium, respectively, is formed inside the groves. The meniscus radius increases as the distance $H$ decreases with increasing bulk helium level, thus gradually shifting the frequency $f_0$ downwards (regime II in Fig.~\ref{fig:3}). Finally, we observe another abrupt drop when the whole device becomes uniformly covered by the superfluid helium (regime III). This behavior is very similar to that observed by employing a superconducting CPW resonator~\cite{YangPRX2016}.    

\begin{figure}[htp]
\centering
\includegraphics[width=1\linewidth]{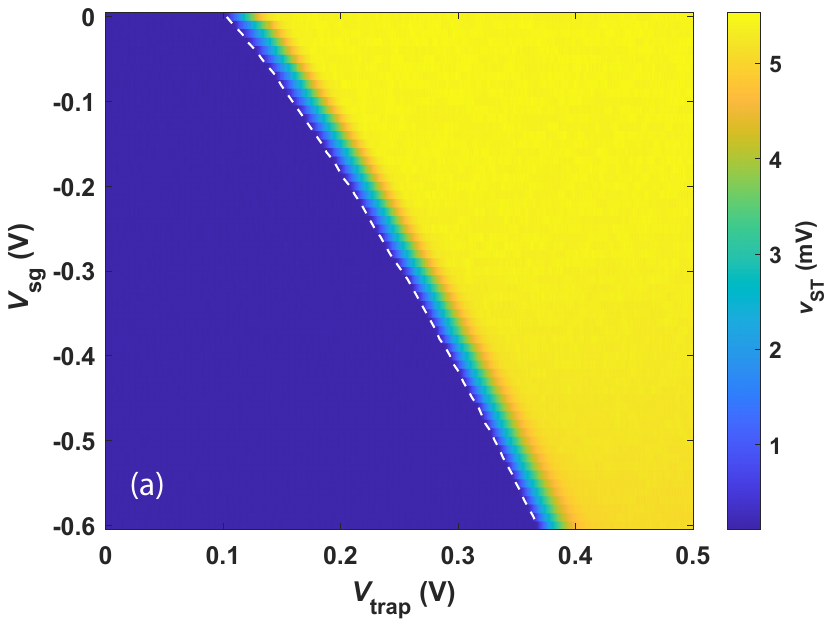}
\includegraphics[width=0.9\linewidth]{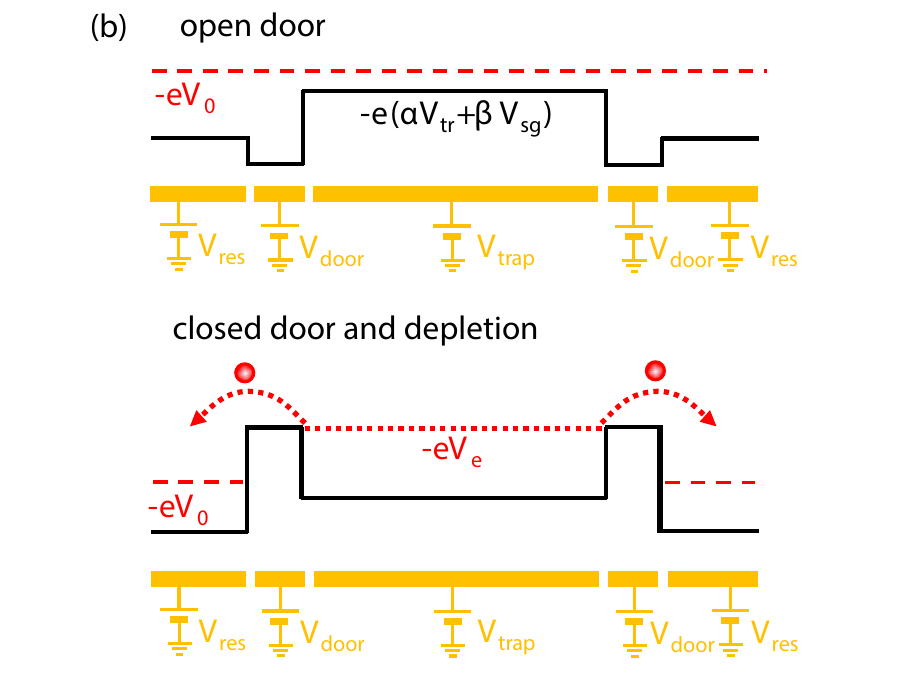}
\caption{\label{fig:4} (color on line) (a) Color map of the Sommer-Tanner signal through the central channel versus the trap bias voltage $V_\textrm{trap}$ (horizontal axis) and spit-gate bias voltage $V_\textrm{sg}$ (vertical axis). The dashed (white) line represents the channel conduction threshold as defined in the text. (b) Simplified representation of the electron potential energy along the central channel (solid line) due to the biased gate electrodes and the electron potential energy in the reservoirs (dashed line) for two voltage configurations (open and closed door) described in the text. The maximum energy of electrons in the trap, at which the electron depletion occurs, is indicated by a dotted line.}
\end{figure}

The above procedure establishes a suitable regime for the confinement and manipulation of electrons in the device. In the experiment described here, the cell is filled with about 4.5~cm$^{3}$ of superfluid helium, and the liquid surface is charged with electrons produced by thermionic emission from a tungsten filament placed about 2~mm above the device. During the charging, a DC bias voltage $V_\textrm{res}$ is applied to the reservoir electrodes and is maintained constant during the measurements. To check the device's FET performance, the standard Sommer-Tanner (ST) method is used~\cite{sommerPRL1971tanner}. An AC driving voltage of 20~mV$_\textrm{rms}$ is applied to one of the reservoir electrodes (source), while an AC current induced by the motion of surface electrons is measured at another reservoir electrode (drain) using a room-temperature transimpedance amplifier (FEMTO DHPCA) followed by a lock-in amplifier. A non-zero ST signal can be observed, providing that there is conduction of electrons between the source and drain through the central channel, which depends on the number of electrons in the channel. The latter can be varied by the DC bias voltages applied to the gate electrodes of the channel. Fig.~\ref{fig:4}(a) shows the color map of the ST signal $v_{\textrm{ST}}$ plotted versus the trap-gate voltage $V_\textrm{trap}$ (horizontal axis) and split-gate voltage $V_\textrm{sg}$ (vertical axis), while a fixed bias voltage $V_\textrm{door}=0.5$~V is applied to the door gate. An abrupt drop of the measured signal to zero indicates the depletion of electrons above the trap gate. The dashed line in Fig.~\ref{fig:4}(a) indicates the conduction threshold defined as the gate voltages at which the ST signal drops below 0.2~mV.

To understand the devices' operation, it is instructive to consider a single-electron potential energy in the channel, see a schematic representation in Fig.~\ref{fig:4}(b). Here, the electron potential energy due to the applied bias voltages is represented by a solid line. The electric potential on the surface of uncharged liquid above the trap electrode is given by $\alpha V_\textrm{trap}+\beta V_\textrm{sg}$, where $\alpha$ and $\beta=1-\alpha$ are the positive weighted contributions to the total capacitance between the surface and the trap and split-gate electrodes, respectively, while the electric potential at the charged surface of liquid in the reservoirs is designated by $V_0$. Because the number of electrons in the reservoirs is very large, $V_0$ can be considered independent of bias voltages applied to the gate electrodes of the central channel. There are two distinct voltage configurations, which we will refer to as the {\it open door} and {\it closed door}. In the open door configuration (see Fig.~\ref{fig:4}(b)), a sufficiently large positive bias $V_{\rm door}$ is applied to the door-gate electrodes to lower electron potential energy above this gate so that electrons are admitted into the channel from the reservoirs. Making a simplifying assumption that electrons are distributed uniformly above the trap gate, forming a parallel-plate capacitor with the electrode, we can establish the following analytical approximation for their areal density~\cite{reesPRL2011pc}

\begin{equation}
n_s=\frac{\epsilon\epsilon_0}{\alpha e d} \left( \alpha V_\textrm{trap} + \beta V_\textrm{sg} - V_0 \right),
\label{eq:2}
\end{equation}          

\noindent where $\epsilon=1.056$ is the dielectric constant of liquid $^4$He and $\epsilon_0=8.85\times 10^{12}$~F/m is the permittivity of free space. The conductance threshold is determined by the condition $\alpha V_\textrm{trap}+\beta V_\textrm{sg}=V_0$, corresponding to $n_s=0$. Above, we assumed that the liquid depth in the channel is equal to the channel height set by the resist thickness $d$. The data shown in Fig.~\ref{fig:4}(a) are taken for the open door configuration. The average slope of the conductance threshold line in this plot gives us a rough estimate of $\alpha/\beta\sim 3$. In this voltage configuration, the density of electrons above the trap electrode can be varied by varying $V_\textrm{trap}$ and $V_\textrm{sg}$ according to Eq.~\eqref{eq:2}. Alternatively, in the closed door configuration, we apply a negative bias voltage $V_{\rm door}$ such that the door region is completely depleted of electrons and a fixed number of electrons above the trap gate is isolated from the electrons in the reservoirs. It is clear that in this case, the electric potential of electrons in the trap $V_{\rm e}$ differs from $V_0$ and varies with the applied voltage $V_{\rm trap}$. However, the electron potential energy $-eV$ can not exceed the potential barrier imposed by the negatively biased door gate because, in this case, some electrons will escape from the trap into the reservoirs. Such a {\it depletion} process is illustrated in Fig.~\ref{fig:4}(b).

\section{Experimental Results}

\subsection{Open door}
\label{sec:open}

The Rydberg transition of electrons above the trap gate is detected by applying mm-wave excitation at a fixed frequency $f_\textrm{mm}$ and tuning the transition frequency of electrons by means the Stark shift in the electric field $E_\perp$ acting on electrons perpendicular to the helium surface. Using the parallel-plate capacitor model employed earlier, the electric field experienced by electrons above the trap gate can be written as

\begin{equation}
E_\perp=\frac{V_\textrm{trap}-V_{\rm e}}{d}-\frac{e n_s}{2\epsilon_0}.
\label{eq:3}
\end{equation}   

\noindent Note that this expression takes into account the contributions to the electric field from both the applied bias voltages and the positive image charge induced by electrons at the gate electrodes. Together with Eq.~\eqref{eq:2}, this allows us to estimate $E_\perp$. For the open door configuration (see Fig.~\ref{fig:4}(b)), we assume a fixed $V_0$ set by the electrons in the reservoirs. Excluding $n_s$ from Eqs.~\eqref{eq:2} and \eqref{eq:3}, we obtain     

\begin{equation}
E_\perp=\frac{V_\textrm{trap}}{d} \left( 1-\frac{\epsilon}{2} \right) - \left( \frac{\beta}{\alpha}\right)\frac{\epsilon V_\textrm{sg}}{2d}+\left( \frac{\epsilon}{2\alpha}-1 \right) \frac{V_0}{d}.
\label{eq:4}
\end{equation}  

\begin{figure}[htp]
\centering
\includegraphics[width=1\linewidth]{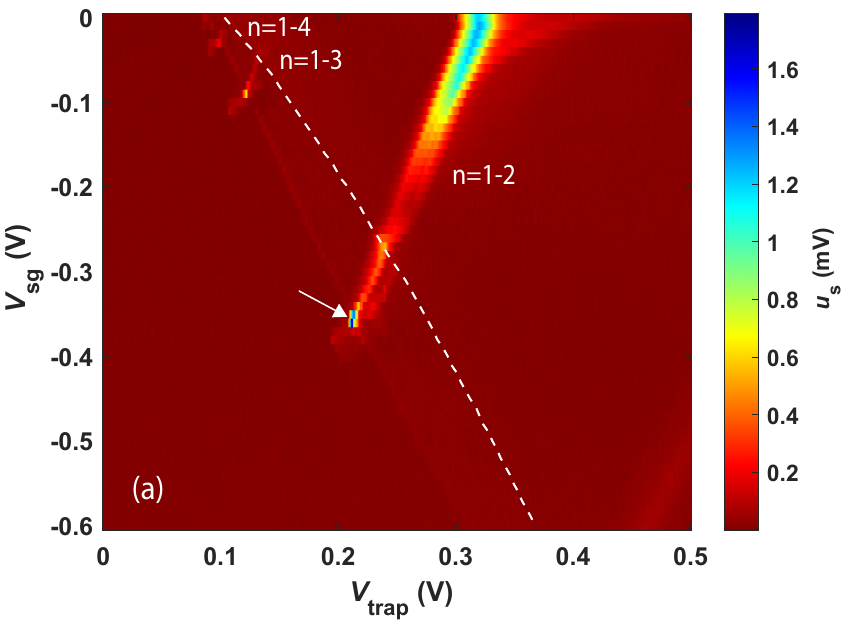}
\includegraphics[width=1\linewidth]{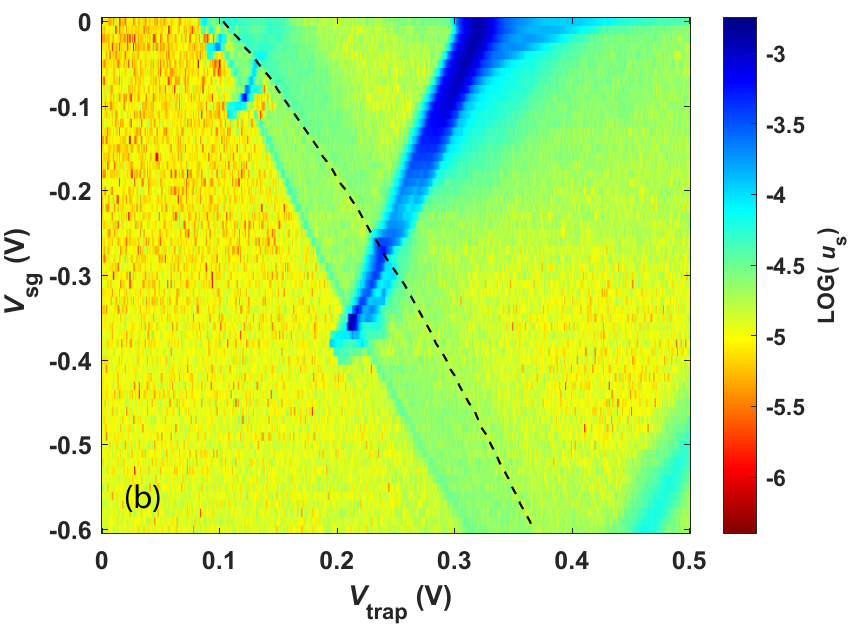}
\caption{\label{fig:5} (color on line) (a) Color map of the image-charge signal $u_{\rm s}$ versus the trap voltage $V_\textrm{trap}$ (horizontal axis) and spit-gate voltage $V_\textrm{sg}$ (vertical axis) obtained with the mm-wave excitation at $f\textrm{mm}=389.58$~GHz, the reservoir bias voltage $V_{\rm res}=0.2$~V, and the door-gate bias voltage $V_\textrm{door}=0.5$~V (open door). An arrow indicates the voltage setting used for the potential energy plot in Fig.~\ref{fig:6}. (c) The log-scale plot of the date shown in panel (b). The dashed in both plots represents the channel conduction threshold as shown in Fig.~\ref{fig:4}(a). All data sets are taken at $T=220$~mK.}
\end{figure}   

In Fig.~\ref{fig:5}(a), we show the image-charge signal measured simultaneously with the ST signal shown in Fig.~\ref{fig:4}(a) and under the mm-wave excitation at frequency $f_\textrm{mm}=389.58$~GHz. The data are collected by varying the trap-gate voltage $V_\textrm{trap}$ from $0.5$~V to zero at a fixed value of $V_\textrm{sg}$ and repeating this procedure for different values of $V_\textrm{sg}$. An image-charge signal due to the Rydberg transition of electrons from the ground $n=1$ state to the first excited $n=2$ state is clearly observed. We also notice that the transitions into higher $n=3$ and 4 excited states are clearly seen and behave similarly. Since the mm-wave excitation frequency is fixed, the gate voltages $V_\textrm{trap}$ and $V_\textrm{sg}$ vary at resonance to ensure the constant value of the pressing field $E_\perp$. According to Eq.~\eqref{eq:4}, the relationship between  $V_\textrm{sg}$ and $V_\textrm{trap}$ on the transition line is expected to be linear with the slope given by $\alpha(2-\epsilon)/(\beta\epsilon)\approx \alpha/\beta$. The dependence observed in Fig.\ref{fig:5}(a) is also linear with a slope equal to approximately 3, which is in reasonable agreement with the estimate of $\alpha/\beta$ from the conduction threshold in Fig.~\ref{fig:4}(a). Unexpectedly, the Rydberg transition is still clearly observed even on the left side of the conduction threshold indicated by the dashed line in Fig.~\ref{fig:5}(a), where according to Fig.~\ref{fig:4}(a) the trap region is supposed to be depleted of electrons. To highlight this intriguing behavior, it is instructive to replot the measured color map in a log scale, as presented in Fig.~\ref{fig:5}(b). It is clear that our image-charge detection system allows us to detect an electron signal not only at the transition resonance of electrons in the trap but also far from the corresponding transition line. In particular, a broad signal on the left of the conduction threshold (dashed) line is clearly distinguishable from the zero-signal background. We believe it originates from the Rydberg transition of the electrons floating above the gaps between the trap and door-gate electrodes, thus experiencing a strong lateral gradient of the vertical electric field $E_\perp$. Because of the high sensitivity of our image-charge detection system, even a small number of such electrons Stark-tuned to the resonance at given bias voltages can be readily detected. Notably, detecting this signal provides us with a valuable tool for device characterization, in addition to the conventional ST method. For example, from Fig.~\ref{fig:5}, we can see that the Rydberg transition line of electrons in the channel extends well beyond the conduction threshold. This indicates that some electrons remain in the trap region even though conduction through the channel has ceased.

\begin{figure}[htp] 
\includegraphics[width=7.5cm,keepaspectratio]{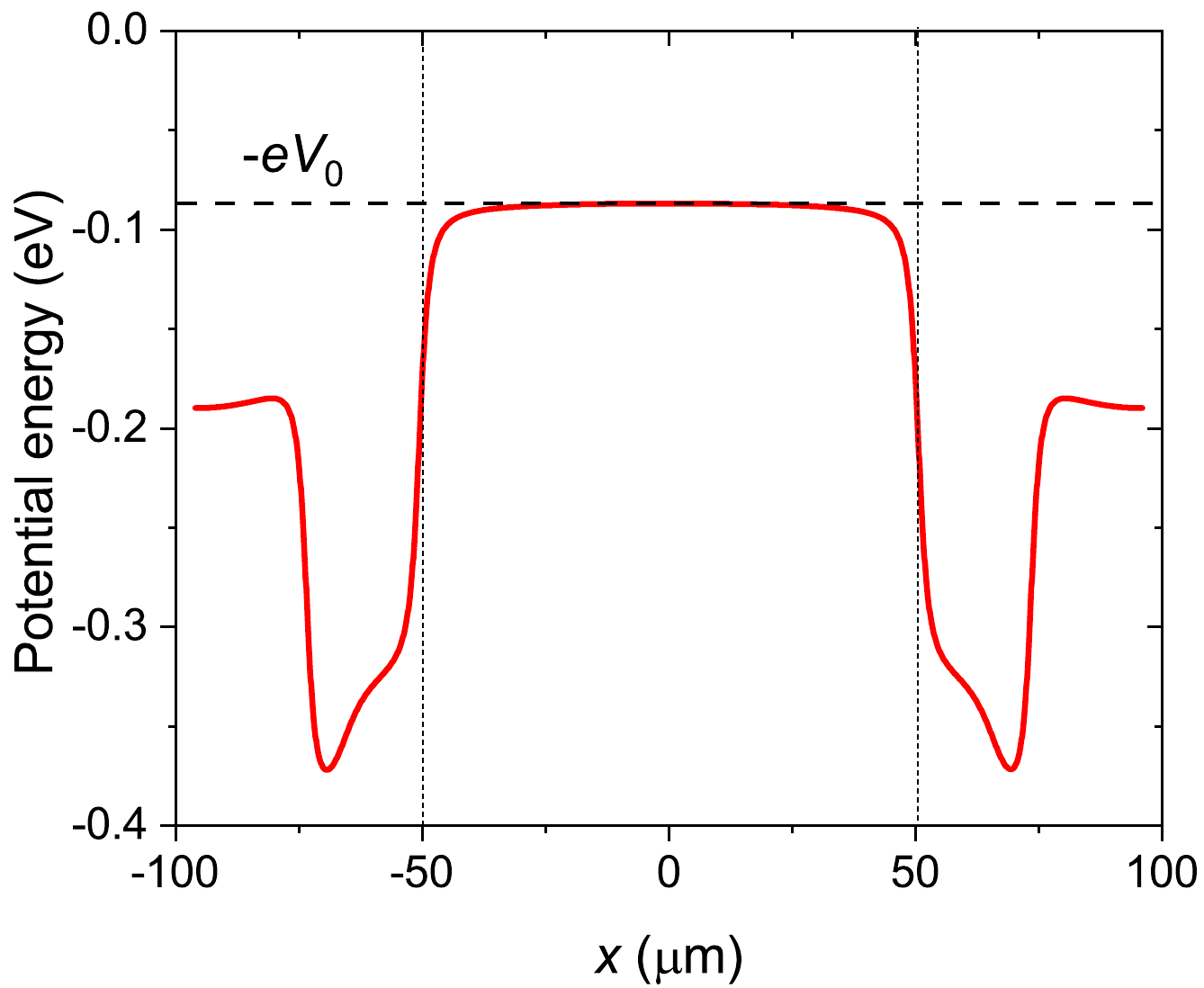} 
\caption{\label{fig:6} (color on line) Calculated potential energy of an electron on the surface of liquid helium along the channel (solid line) due to the applied gate voltages $V_\textrm{door}=0.5$~V, $V_\textrm{trap}=0.21$~V and $V_\textrm{sg}=-0.36$~V corresponding to the point indicated by an arrow in Fig.~\ref{fig:5}(a). The dashed line represents the potential energy of electron on the charged surface above the reservoirs equal to $-eV_0$.}
\end{figure}   

To understand this unexpected behavior, one should consider beyond an oversimplified model of the electron potential energy represented by the solid line in Fig.~\ref{fig:4}(b). To do that, we carried out a numerical FEM simulation to calculate the distribution of the electric potential on the surface of the liquid in the channel due to applied gate voltages. In the simulation, we choose the $xy$ plane to be along the liquid helium surface, with the $x$-axis directed along the channel and the origin chosen at the middle of the channel. An example of simulation for $V_\textrm{door}=0.5$~V, $V_\textrm{trap}=0.21$~V and $V_\textrm{sg}=-0.36$~V is shown in Fig.~\ref{fig:6} by the solid (red) line corresponding to $y=0$ (center of the channel). Note that the gate voltages correspond to the point marked by an arrow in Fig.~\ref{fig:5}(a). The horizontal dashed line represents the potential energy of an electron $-eV_0$ on the charged surface above the reservoirs, while the short-dashed lines indicate the position of gaps between the trap and door electrodes. From this calculation it is clear that due to the convex shape of the electric potential along the channel, which arises due to the finite length of the trap electrode, the electron density depletes non-uniformly along the channel as $-eV_0$ intersects the maximum of the solid line. In particular, when the electron density is zero at the center of the channel, and the conduction through the channel ceases, there are still some electrons on both sides of the trap region which contribute to the image-charge signal observed in Fig.~\ref{fig:5}(a,b).  

\begin{figure}[htp] 
\includegraphics[width=\columnwidth,keepaspectratio]{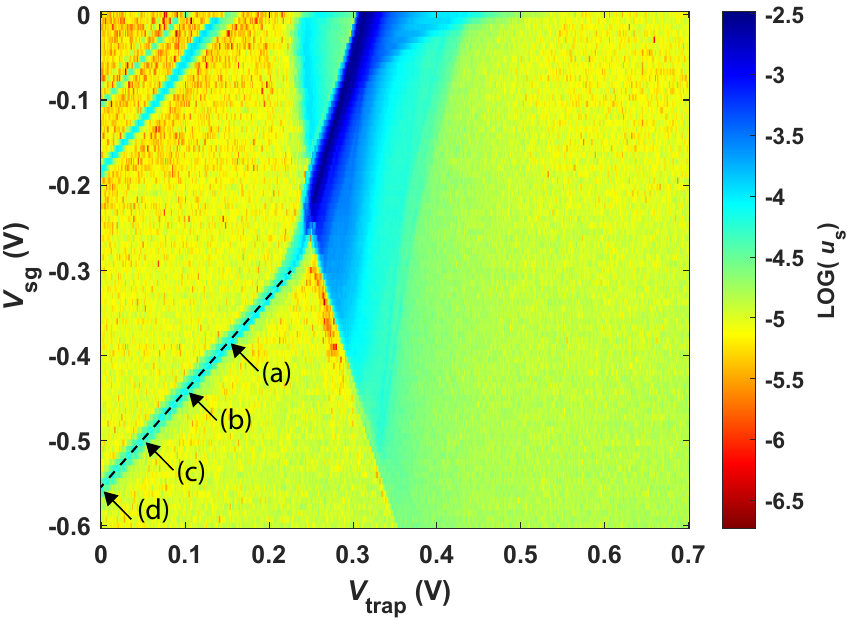} 
\caption{\label{fig:7} (color on line)  Log-scale color map of the measured image-charge signal versus the bias voltages $V_\textrm{trap}$ and $V_\textrm{sg}$ (vertical axis) obtained with the mm-wave excitation at $f\textrm{mm}=389.58$~GHz and the door-gate bias voltage $V_\textrm{door}=V_\textrm{trap}$. The dashed line represents a straight line with the slope equal 1. The arrows indicate voltage settings used for the potential energy plots in Fig.~\ref{fig:8}.}
\end{figure}

It is instructive to carry out another measurement where the procedure is similar to that used to obtain Fig.~\ref{fig:5}(a,b), but the trap and door gate voltages are always maintained equal, $V_\textrm{trap}=V_\textrm{door}$. The results are shown in Fig.~\ref{fig:7} where the measured image-charge signal $u_\textrm{s}$ is plotted in the log scale. As in the log-scale plot in Fig.~\ref{fig:5}(b), a broad transition line is observed, indicating the presence of electrons in the trap region of the device. Surprisingly, narrow transition lines corresponding to the first and higher excited states are clearly observed for the gate voltages where the central channel is supposed to be completely depleted of electrons. The lines have a linear dependence on gate voltages, with the slope being very close to 1. These transition lines are found to extend down to large negative values of $V_\textrm{trap}$ without significant change in slope.

\begin{figure}[htp] 
\includegraphics[width=7.5cm,keepaspectratio]{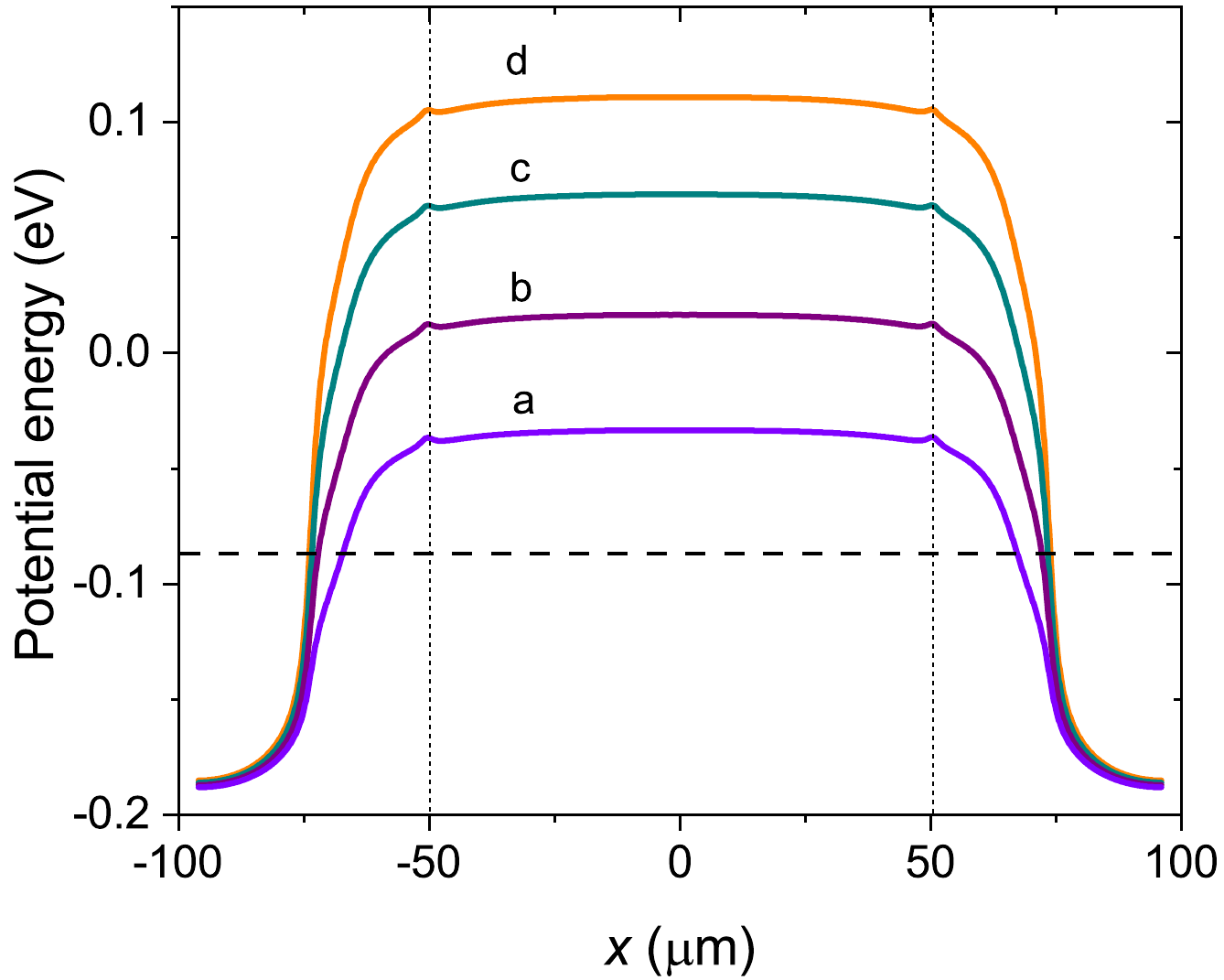} 
\caption{\label{fig:8} (color on line)  Potential energy of an electron on the surface of liquid helium along the channel (solid lines) for the applied gate-bias voltages (a) $V_\textrm{trap}=V_\textrm{door}=0.15$~V and $V_\textrm{sg}=-0.39$~V, (b) $V_\textrm{trap}=V_\textrm{dr}=0.1$~V and $V_\textrm{sg}=-0.44$~V, (c) $V_\textrm{tr}=V_\textrm{dr}=0.05$~V and $V_\textrm{sg}=-0.5$~V, and (d) $V_\textrm{tr}=V_\textrm{dr}=0.01$~V and $V_\textrm{sg}=-0.55$~V. The dash line represents the potential energy of electron $-eV_0$ as in Fig.~\ref{fig:6}.}
\end{figure}

As previously, we can account for this behavior by considering the electric potential distribution on the surface along the channel due to the applied gate voltages. Fig.~\ref{fig:8} shows $y=0$~\textmu m slices of the calculated electron potential energy profile on the helium surface for several sets of gate bias voltages marked by arrows in Fig.~\ref{fig:7}. Note that for all gate voltages considered here, the electron potential energy $-eV_0$ (indicated by the horizontal dashed line) is much lower than the electron potential energy above the trap gate. Therefore, the trap region is expected to be completely depleted of electrons. A prominent feature of the energy plots is the presence of shallow potential minima near the insulating gaps between the trap and door electrodes, whose position is indicated by vertical short-dashed lines. Such shallow minima can trap a certain number of electrons. The Rydberg resonance of such trapped electrons is detected in our experiment thanks to the high sensitivity of the image-charge detection system. As shown later, the voltage dependence of the transition line for such electrons is expected to be linear with the slope equal to 1, in agreement with the result in Fig.~\ref{fig:7}.            

\subsection{Closed door and depletion} 
\label{sec:closed}

In the closed-door voltage configuration, with a negative bias $V_{\rm door}$ applied to the door electrode, a fixed number of electrons can be confined and manipulated in the trap region, see Fig.~\ref{fig:4}(b). By excluding the electron potential energy of such electrons from Eqs.~\eqref{eq:2} and \eqref{eq:3}, we obtain the following expression for the electric field $E_\perp$ experienced by such electrons

\begin{equation}
E_\perp=\frac{\beta\left( V_\textrm{trap}-V_\textrm{sg}\right)}{d} + \left(2\alpha - \epsilon\right) \frac{en_s}{2\epsilon\epsilon_0}.
\label{eq:5}
\end{equation}  

\begin{figure}[htp] 
\includegraphics[width=\columnwidth,keepaspectratio]{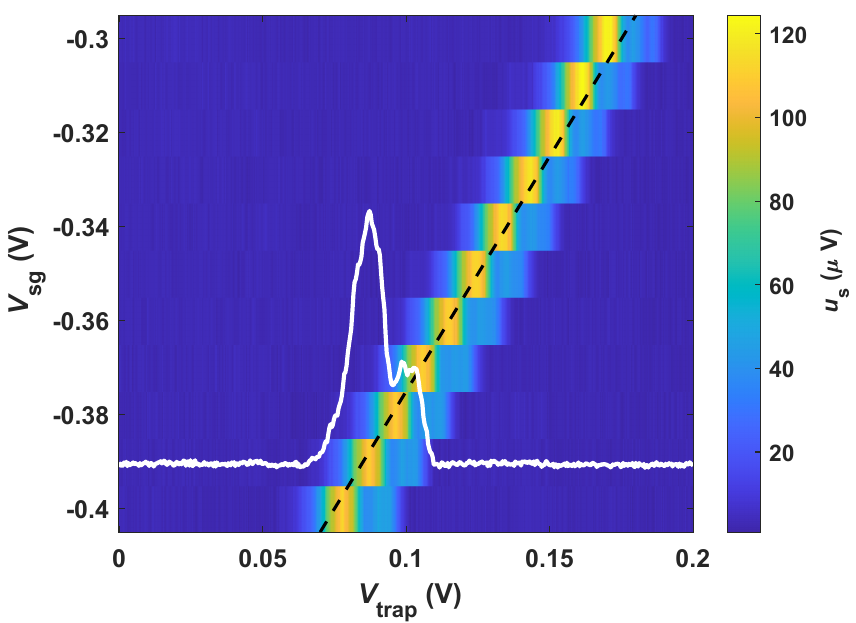} 
\caption{\label{fig:9} (color on line)  Color map of the image-charge signal versus the trap voltages $V_{\rm trap}$ and $V_{\rm sg}$ obtained with the mm-wave excitation at $f{\rm mm}=416.58$~GHz, the reservoir bias voltage $V_{\rm res}=0.5$~V, and the door-gate bias voltage $V_{\rm door}=-0.4$~V (closed door). The dashed line represents a straight line with slope equal 1. A representative Stark spectrum $u_{\rm s}$ versus $V_{\rm trap}$ obtained for $V_{\rm sg}=-0.39$~V is given by a solid line.}
\end{figure}

\noindent If we assume that the charge density of the confined electrons does not change significantly by varying the gate voltages, according to the above equation, we expect a linear dependence between $V_{\rm sg}$ and $V_{\rm trap}$ on the transition line, with the slope equal to 1. Fig.~\ref{fig:9} shows the image-charge signal versus $V_{\rm trap}$ and $V_{\rm sg}$ measured with the mm-wave excitation at $f_{\rm mm}=416.58$~GHz and a fixed door voltage $V_{\rm door}=-0.4$~V. The Rydberg transition line is observed, with the linear voltage dependence represented by a dashed line. The slope of this line is equal to 1, in excellent agreement with Eq.~\eqref{eq:5}. A representative Stark spectrum $u_{\rm s}$ versus $V_{\rm trap}$ obtained at a fixed value of $V_{\rm sg}=-0.39$~V is plotted by a solid line. The spectrum's shape points out the inhomogeneous broadening of the transition line in a non-uniform electric field $E_\perp$ inside the channel, as will be discussed later.       

\begin{figure}[htp] 
\includegraphics[width=\columnwidth,keepaspectratio]{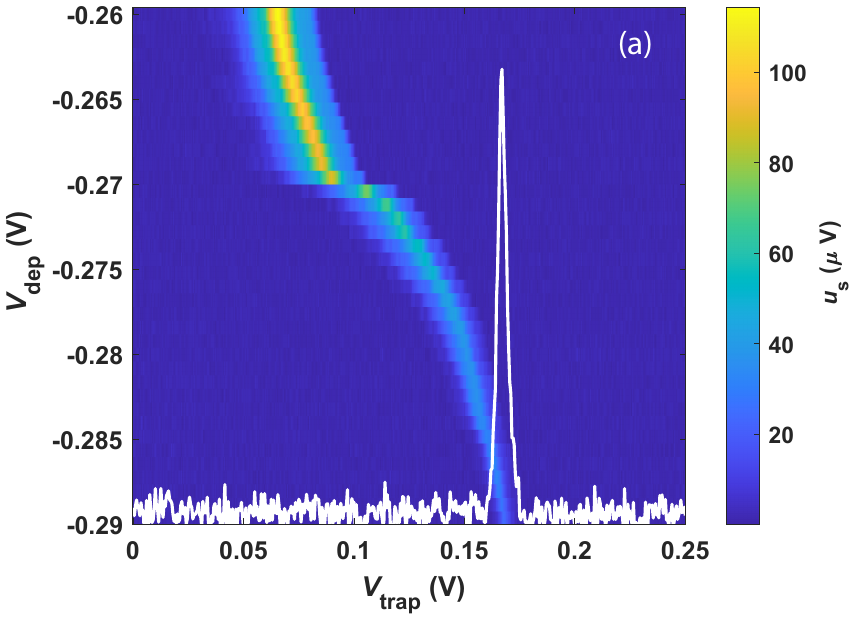} 
\includegraphics[width=7.5cm,keepaspectratio]{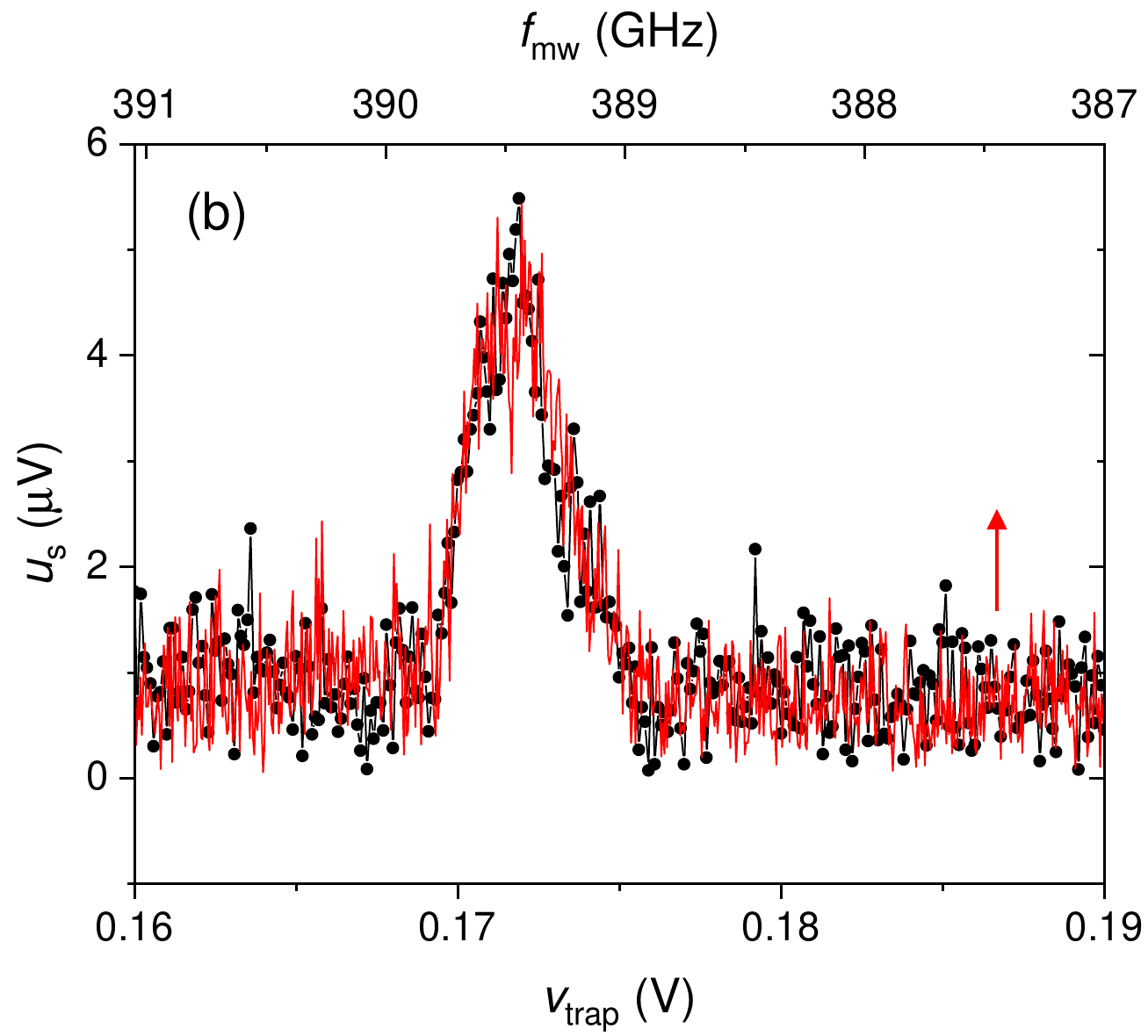} 
\caption{\label{fig:10} (color on line) (a) Color map of the measured image-charge signal versus the trap voltage $V_{\rm trap}$ for different depletion voltages $V_{\rm dep}$. The signal is measured with the mm-wave excitation at $f_{\rm mm}=389.52$~GHz, the reservoir bias voltage $V_{\rm res}=0.5$~V, the door bias voltage $V_{\rm door}=-0.4$~V (closed door) and the split-gate bias voltage $V_{\rm sg}=-0.4$~V . A representative Stark spectrum $u_{\rm s}$ versus $V_{\rm trap}$ obtained for $V_{\rm dep}=-0.29$~V is given by a solid line. (b) The image-charge signal measured at a fixed mm-wave frequency $f_{\rm mm}=389.52$~GHz by varying $V_{\rm trap}$ (solid cycles) and fixed trap voltage $V_{\rm trap}=0.1719$~V by varying $f_{\rm mm}$ (solid line) for the depletion voltage $V_{\rm dep}=-0.2945$~V.}
\end{figure}  

The closed door configuration of gate voltages suggests the following {\it depletion procedure} to control the number of electrons isolated in the trap. Initially, the channel is filled with electrons from the reservoirs by applying a large positive voltage to both trap and door electrodes. Then, the door voltage $V_{\rm door}$ is set to a negative value to isolate an initial number of electrons in the trap. After that, the trap voltage $V_{\rm trap}$ is decreased to a certain value $V_{\rm dep}$, which we will call the depletion voltage. Correspondingly, the potential energy of electrons in the trap $-eV_{\rm e}$ increases. When it exceeds the potential barrier imposed by the negatively biased door gate, some electrons escape from the trap into the reservoirs until $V_{\rm e}$ is equal to the electric potential of the uncharged surface of liquid above the door gate (see Fig.~\ref{fig:4}(b)). After the depletion, the trap voltage $V_{\rm trap}$ is returned to the initial value, and a voltage scan can be performed to observe the image-charge signal due to the Rydberg transition. Note that the split-gate voltage $V_{\rm sg}$ is maintained constant throughout the entire procedure. A color map of the measured image-charge signal versus $V_{\rm trap}$ for different values of the depletion voltage $V_{\rm dep}$ is shown in Fig.~\ref{fig:10}(a). The data are taken at a fixed door voltage $V_{\rm door}=-0.4$~V and split-gate voltage $V_{\rm sg}=-0.4$~V. The number of electrons in the trap decreases with decreasing depletion voltage $V_{\rm dep}$. Correspondingly, the image-charge signal due to the Rydberg transition decreases in amplitude and the resonance shifts towards the higher value of $V_{\rm trap}$. This can be easily understood from Eq.~\eqref{eq:5}. For an estimated value of $\alpha \approx 0.7$, the second term on the left-hand side gives a positive contribution to $E_\perp$. As the number of electrons decreases, the first term must be increased by increasing $V_{\rm trap}$ to maintain the constant value of $E_\perp$. The transition line also becomes narrower with a decreasing number of electrons. Such narrowing is expected from the inhomogeneous broadening due to a nonuniform electric field $E_\perp$ in the trap since a smaller number of electrons occupy a smaller area in the trap. A representative Stark spectrum obtained for the depletion voltage $V_{\rm dep}=-0.29$~V is plotted by a solid line in Fig.~\ref{fig:10}(a). It is observed that the transition line shape becomes more homogeneous for a decreasing number of electrons. In order to reliably establish the relationship between the Rydberg transition frequency $f_{21}$ and the applied bias voltage $V_{\rm trap}$, we performed an independent measurement of the transition line by sweeping the mm-wave frequency $f_{\rm mm}$ at a fixed voltage $V_{\rm trap}$ and compared it with the Stark spectrum taken at a fixed mm-wave frequency. The comparison is shown in Fig.~\ref{fig:10}(b) where the image-charge signals measured at a fixed mm-wave frequency $f_{\rm mm}=389.52$~GHz and a fixed trap voltage $V_{\rm trap}=0.1719$~V are plotted by the solid (black) cycles and solid (red) line, respectively. Here, the frequency range on the upper horizontal axis is adjusted to overlap two spectra. It is found that the line shape of the two spectra matches very well, suggesting a linear relationship between $f_{\rm mm}$ and $V_{\rm trap}$ with the slope equal to approximately 0.13~GHz/mV.  

\begin{figure}[htp] 
\includegraphics[width=7.5cm,keepaspectratio]{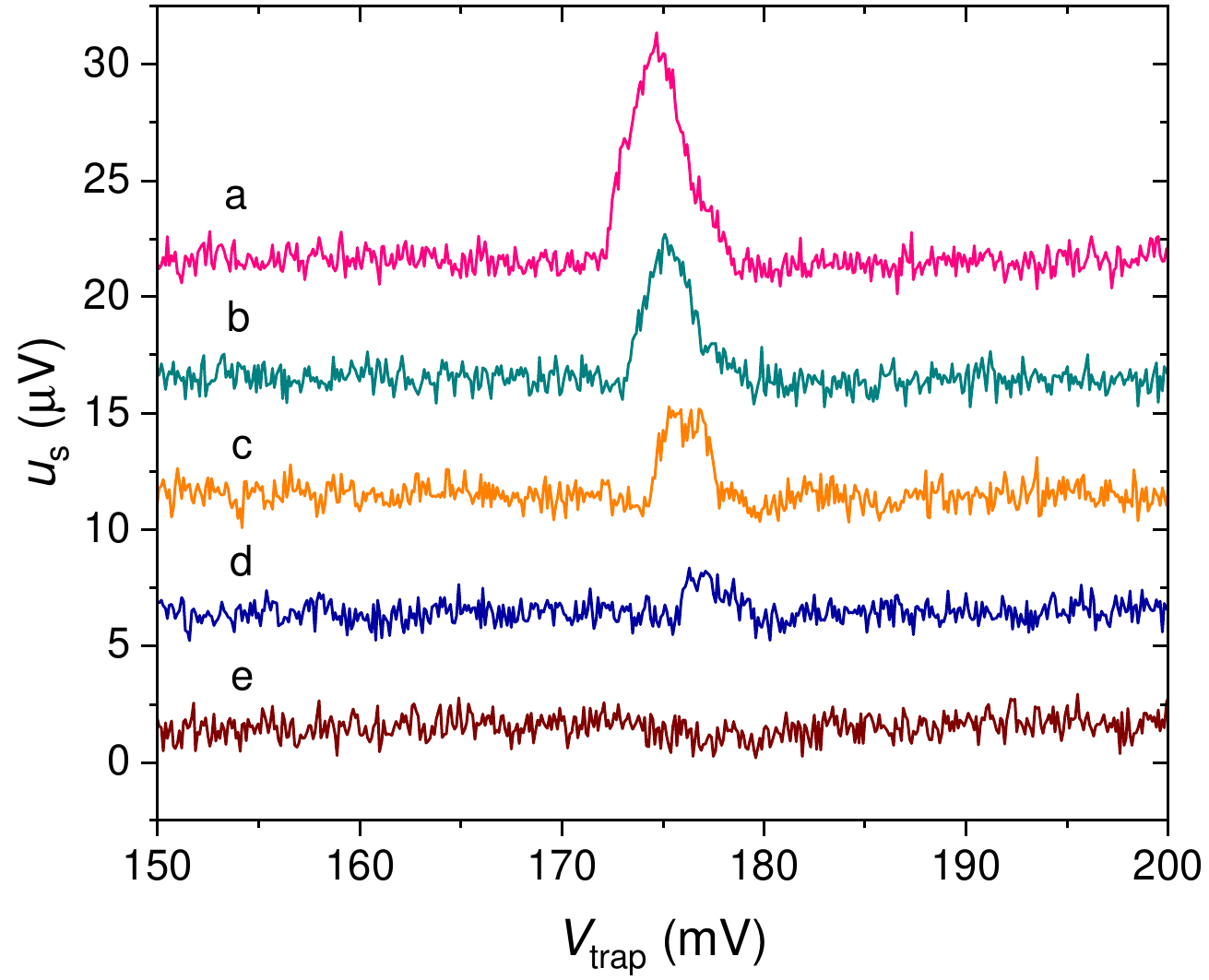} 
\caption{\label{fig:11} (color on line) Representative Stark spectra of the electrons in the trap for several different value of the depletion voltage $V_{\rm dep}=-0.2943$ (a), -0.2954 (b), -0.2969 (c), -0.2983~V (d), and -0.2994 measured with the mm-wave excitation frequency $f_{\rm mm}=389.52$~GHz. The (a), (b), (c) and (d) traces are shifted by 20, 15, 10 and 5~$\mu$V, respectively, for clarity.}
\end{figure}      

Finally, in order to test the sensitivity of our image-charge detection system, we measure the Rydberg resonance by decreasing the number of electrons in the trap until the transition signal is lost. Some exemplary Stark spectra taken at $V_{\rm dep}=-0.2943$, -0.2954, -0.2969, -0.2983, and -0.2994~V are shown in Fig.~\ref{fig:11}. The other conditions are the same as for the data shown in Figs.~\ref{fig:10}(a,b). The bandwidth for this measurement is set at about 0.2~Hz (the lock-in settling time of 2 s). The signal is not observed for the depletion voltage below -0.2983~V, either because it is buried in noise or the trap is completely depleted of electrons. Note that the RMS noise voltage of approximately 0.3~$\mu V$ observed for the trace (e) in Fig.~\ref{fig:11} corresponds to $0.55$~nV/$\sqrt{\textrm{Hz}}$ of the noise voltage at the input of an ideal amplifier (voltage gain $A=1,200$). This is in excellent agreement with our earlier estimation of the noise spectral density in Fig.~\ref{fig:2} of Sect.~\ref{sec:noise}. The peak value $u_s=2$~$\mu$V of the measured signal at $V_\textrm{dep}=-0.2983$~V, see the trace (d) in Fig.~\ref{fig:11}, corresponds to an image current induced at the trap electrode $i_s=u_s/(A\kappa R)\approx 2.8$~fA, with $R\approx 18$~$\textrm{M}\Omega$ (see Sect.~\ref{sec:noise}). This current is approximately equal to an estimated current signal of $2.9$~fA from a single electron excited with the probability $P_{\rm e}=1$. For a more realistic estimation, we have to consider the mean occupation number for the first excited state. For the stationary state, this can be calculated as $\rho_{22}=\omega_1^2/(2\omega_1^2+\gamma^2)$ from the optical Bloch equations, where $\omega_1$ is the Rabi frequency and $\gamma$ is the intrinsic linewidth. At the temperature $T= 220$~mK, we calculate $\gamma/2\pi=10$~MHz due to the quasi-elastic scattering of an electron from ripplons~\cite{andoJPSJ1978}. The Rabi frequency is given by $\omega_1=ez_{12}E_\textrm{mm}/\hbar$, where $z_{12}\approx 3.6$~nm is the transition moment and $E_\textrm{mm}$ is the mm-wave electric field. The power of mm-wave radiation at $f_\textrm{mm}=389.58$~GHz emitted from an open end of WR-6 waveguide inserted into the experimental cell was measured at the room temperature to be about $P=40$~$\mu$W. Using the specified 6.5~dBi directivity of the WR-6 waveguide, we expect the electric field $E_\textrm{mm}=10^{6.5/20}\sqrt{P\eta_0/(2\pi R^2)}\approx 5.7$~V/m ($\eta_0=377$~Ohm is the wave impedance in vacuum) at a distance $R=18$~mm between an open end of the waveguide and the on-chip device. This corresponds to the Rabi frequency $\omega_1/2\pi=5$~MHz and the mean occupation number $\rho_{22} \approx 0.2$. From this estimate, the peak amplitude of the trace (d) in Fig.~\ref{fig:11} corresponds to about 5 excited electrons at a fixed value of $V_{\rm trap}$. However, we also find that the signal is significantly broadened. The half-width of the trace (d) in Fig.~\ref{fig:11} is approximately 1.5~mV. Using the level arm between the Rydberg transition frequency $f_{21}$ and the trap-gate voltage $V_\textrm{trap}$ of 0.13~GHz/mV determined earlier, we estimate the transition line broadening $\Delta f_{21}\approx 200$~MHz. This is significantly larger than the intrinsic transition linewidth $\gamma/2\pi=10$~MHz, which suggests that the total signal comes from many electrons spread over a large area of the trap and inhomogeneously broadened by a nonuniform electric field $E_\perp$, as was discussed earlier. The total number of electrons contributing to the observed signal scales as a ratio $\Delta f_{21}/\gamma~\approx 20$, thus we can estimate the total number of electrons contributing to a line shape of the signal (d) in Fig.~\ref{fig:11} to be approximately one hundred.

We can also estimate the number of electrons lost from the trap during depletion by considering the voltage shift of the transition spectra in Fig.~\ref{fig:11} with decreasing number of electrons. According to Eq.~\ref{eq:5}, the resonance shifts towards higher values of $V_{\rm trap}$ to compensate for the decrease in the electron density $n_s$. From Eq.~\ref{eq:5}, the shift of the transition resonance $\Delta V_{\rm trap}\approx 1$~mV observed in Fig.~\ref{fig:11} corresponds to a change in the electron density $\Delta n_s \approx \epsilon\epsilon_0\Delta V_{\rm trap}/(ed)=0.04$~$\mu{\rm m}^{-2}$. Taking the area of the trap $10\times 100=10^3$~$\mu{\rm m}^2$, this corresponds to about 40 electrons lost from the trap during each depletion cycle. Such an estimate supports the conclusion that the depletion following the acquisition of trace (d) in Fig.~\ref{fig:11} empties electrons from the trap, such that no transition signal can be observed in trace (e) in Fig.~\ref{fig:11}.  

\section{Discussion and Conclusions}

In summary, we have demonstrated the detection of the Rydberg transition in a small ensemble of electrons confined and manipulated in a single helium-filled microchannel of an electron-on-helium FET device. A three-electrode gate structure allowed us to trap a controllable number of electrons in the channel while the image-charge detection system was used to record their Stark-tuned transition spectra. Thanks to the very high sensitivity of our detection system based on a superconducting helical resonator, we were able to observe the transition signal from a few excited electrons, with the total number of electrons in the trap contributing to the inhomogeneous broadening of the resonance to be on the order of hundreds. However, we note that, in the above estimation, we have assumed that the inhomogeneous broadening comes exclusively from a nonuniform clamping electric field $E_\perp$ exerted on the electrons. One should not exclude other mechanisms that could lead to the broadening of the single-electron transition. For example, it has been shown that fluctuations of the helium surface induced by the mechanical vibration of the cryostat can lead to dephasing of the electronic states of lateral motion and significant broadening of the corresponding transition line~\cite{Koolstra2019}. In our case, according to Eqs.~\eqref{eq:3} fluctuations in the liquid depth $d$ would lead to fluctuations in the Stark shift of the Rydberg levels and corresponding transition frequencies, thus producing an additional line broadening. In particular, an RMS noise in the helium depth of the order 1~nm due to the mechanical vibrations was found in the microchannel devices with similar geometry as ours~\cite{Koolstra2019,beysJLTP2022}. In our setup, this will produce a broadening of the Rydberg transition on the order 50~MHz, which is significantly larger than the expected intrinsic linewidth and would contribute to the broadening observed here. Since, for such a mechanism, the same electrons contribute to the broadening of the transition line, the actual number of electrons in the trap will be smaller than the above estimation.

While the sensitivity of our detection system is sufficient to observe the transition signal due to a single electron, providing the microwave excitation rate is sufficiently high, the sources of extrinsic broadening have to be eliminated as much as possible because the line broadening decreases the signal amplitude. Ideally, an electron's lateral motion must be quantized in a single-electron trap to eliminate the effect of inhomogeneous clamping field $E_\perp$~\cite{kawakamiPRA2023qubits}. In addition, the quantization of the lateral motion suppresses the elastic scattering of electrons from a single ripplon, thus limiting the relaxation rate of the excited Rydberg state to about 1~MHz due to the spontaneous emission of two short-wavelength ripplons~\cite{kawakamiPRA2023qubits,monarkhaFNT2010decay}. The quantization of the electronic lateral motion can be accomplished by reducing the size of the trap electrode to $\sim 1$~$\mu$m. A split-gate electrode structure instead of door electrodes could be helpful for a controllable loading of such a trap with electrons, as recently demonstrated~\cite{cast2024}. Also, the image-charge detection electrode has to be well decoupled from the stray electrons outside the trap because, as shown in the present work, they can contribute to the measured image-charge signal. Other sources of line broadening and decoherence, such as helium surface fluctuations discussed earlier, must be reduced as much as possible.

Further improvement in the sensitivity of the detection system is achievable, in particular by increasing the quality factor $Q$ of the helical resonator. The strong temperature dependence of $Q$ below 1 K (see Fig.~\ref{fig:2app}) indicates that it is not limited by losses in the superconducting NbTi, which has the transition temperature of about 10~K, but rather losses in a soldered wire joint used to attach the coil to the experimental cell. It was shown that a loaded quality factor exceeding 10,000 can be achieved for such resonators even under strong magnetic fields of several Tesla~\cite{ulmRSI2009,ulmNuc2013,nagaRSI2016}. Increasing $Q$ enhances the resonant impedance of the tank circuit $R=2\pi f_0 LQ$, and therefore S/N ratio $\sim\sqrt{R}$, as shown in Sec.~\ref{sec:noise}. The ability to work in strong magnetic fields exceeding 1 T is important for the proposed non-destructive detection of the spin state of a trapped electron for quantum logic gate implementation~\cite{kawakamiPRA2023qubits}.

Another improvement in the experimental setup can be made by enhancing the Rabi frequency of the Rydberg transition. In the present setup, the mm-wave radiation is emitted from an open end of an overmoded WR-6 waveguide terminated inside the cell at 18 mm from the sample (see Fig.~\ref{fig:1}). Straightforward increasing of the input mm-wave power is not allowed because the emitted radiation, which is mainly absorbed by the inner wall of the experimental cell, overheats the liquid helium and significantly increases the rate of electron scattering from ripplons and helium vapor atoms~~\cite{andoJPSJ1978}. A possible solution to increasing the excitation rate is quasi-optical focusing of mm-wave radiation onto the trapped electrons~\cite{goldsmith1998quasioptical}. In this approach, the mm-wave radiation propagating in free space is focused by a collimating high-resistivity silicon lens, which can simultaneously serve as a substrate for the on-chip trapping device~\cite{Buttgenbach1993AnIS}. We can further enhance electric field strength by integrating trap electrodes with a planar broadband antenna and engineer electric field polarization for efficient radiation coupling with the Rydberg transition~\cite{GONZALEZ2005418}.

The combination of single-electron traps with conducting channels packaged in an on-chip microfluidic device offers the possibility for a modular approach toward the scalability of the quantum logic gates. Such an approach is used for the proposed large-scale quantum computers based on trapped electrons and ions, where the quantum operations are limited to multiple tiny crystals that can be rearranged by splitting and shuttling operations~\cite{kielpinski2002,molmSA2017,Pino2021,haffPRA2022}. A somewhat similar approach is undertaken for a large-scale semiconductor quantum computer, where shuttling spins across a chip between quantum dots might provide a viable and CMOS-compatible way towards scalability~\cite{fujNJP2017,milNatComm2019,yonNatComm2021,zwePRX2023,lanPRX2023}. The mobile electrons on superfluid helium present an excellent candidate to realize a quantum charge-coupled device (QCCD) architecture, owing to the long coherence of the spin state, insensitivity to charge noise thanks to a weak intrinsic spin-orbit interaction, and exceptionally high shuttling fidelity using a simple 3-phase CCD clock sequence~\cite{Lyon2006,bradburyPRL20011ccd}. Such operations can be used to bring a selected pair of electrons from dedicated storage regions into the interaction region, where a localized magnetic field gradient is used to entangle their spin states~\cite{kawakamiPRA2023qubits}, then moved back to storage or a spin-state readout region. The realization of such architecture is the primary motivation for our current and future work.                       

{\bf Acknowledgements} This work is supported by the internal grant from the Okinawa Institute of Science and Technology (OIST) Graduate University and the Grant-in-Aid for Scientific Research (Grant No. 23H01795 and 23K26488) KAKENHI MEXT.

$   $
\providecommand{\noopsort}[1]{}\providecommand{\singleletter}[1]{#1}%

\end{document}